\documentclass[letterpaper, 10pt, conference, onecolumn]{ieeeconf}

\IEEEoverridecommandlockouts                              
\usepackage{mathtools}

\DeclarePairedDelimiter\floor{\lfloor}{\rfloor}

\title{On Distributed Internal Model Principle for Output Regulation over \\ Time-Varying Networks of Linear Heterogeneous Agents}

\author{Kai Cai \ \ \ (\today) 
\thanks{K. Cai is with Urban Research Plaza, Osaka City University, Japan. %
Email: kai.cai@eng.osaka-cu.ac.jp. This work was supported in part
by JSPS KAKENHI Grant Number 26870169 and Program to Disseminate Tenure Tracking System, MEXT, Japan.}
}

\usepackage{graphicx}
\usepackage{amssymb,amscd,amssymb,amsmath,mathrsfs}
\usepackage[sort,compress]{cite}
\usepackage[noend]{algpseudocode}
\usepackage{multirow}
\usepackage{color}
\usepackage{url}

\newtheorem{thm}{Theorem}
\newtheorem{lem}{Lemma}

\begin{document}

\maketitle \thispagestyle{titlepagestyle} \pagestyle{empty}

\begin{abstract}
We study a multi-agent output regulation problem, where not all agents have access to the exosystem's dynamics. We propose a distributed controller that solves the problem for linear, heterogeneous, and uncertain agent dynamics as well as time-varying directed networks. The distributed controller consists of two parts: (1) an {\it exosystem generator} that creates a local copy of the exosystem dynamics by using consensus protocols, and (2) a {\it dynamic compensator} that uses (again) consensus to approach the internal model of the exosystem and thereby achieves perfect output regulation.  Our approach leverages methods from internal model based controller synthesis, multi-agent consensus over directed networks, and stability of time-varying linear systems; the derived result is an adaptation of the (centralized) internal model principle to the distributed, networked setting.
\end{abstract}

\section{Introduction}
\label{Sec_Intro}

Over the past decade, many distributed control problems of networked multi-agent systems have been extensively studied; these include e.g. consensus, averaging, synchronization, coverage, and formation (e.g. \cite{JadLinMor:03, OlfMur:04, BulCorMar:09, CaiIshii:AUT12, LinWanHanFu:14}. Progressing beyond the first/second-order and homogeneous agent dynamics, the {\it distributed output regulation problem} with general linear (time-invariant, finite-dimensional) and heterogeneous agent dynamics has received much recent attention \cite{XiaWeiLi:09,WanHonHuaJia:10,SuHua:TAC12,SuHonHua:13}. In this problem, a network of agents each tries to match its output with a reference signal, under the constraint that only a few agents can measure the reference directly. The reference signal itself is typically generated by an external (linear) dynamic system, called ``exosystem''.  The distributed output regulation problem not only subsumes some earlier problems like consensus and synchronization (viewing the exosystem as the ``virtual leader''), but also addresses issues of disturbance rejection and robustness to parameter uncertainty.

Output regulation has a well-studied centralized version: A single plant tries to match its output with a reference signal (while maintaining the plant's internal stability) \cite{Dav:76,FraWon:75,FraWon:76,Fra:77}. In the absence of system parameter uncertainty, the solution of the ``regulator equations'', embedding a copy of the exosystem dynamics, provides a solution to output regulation \cite{Fra:77}.  When system parameters are subject to uncertainty, a dynamic compensator/controller must be used embedding $q$-copy of the exosystem, where $q$ is the number of (independent) output variables to be regulated. The latter is well-known as the {\it internal model principle} \cite{FraWon:76}.  These methods for solving the centralized output regulation problem, however, cannot be applied directly to the distributed version, inasmuch as not all agents have access to the reference signal or the exosystem dynamics.

The distributed output regulation of networks of heterogeneous linear agents is studied in \cite{SuHua:TAC12}.  The proposed distributed controller consists of two parts: an exosystem generator and a controller based on regulator equation solutions. Specifically, the exosystem generator of each agent aims to (asymptotically) synchronize with the exosystem using consensus protocols, thereby creating a local copy of the exosystem. Meanwhile each agent independently tracks the signal of its local generator, by applying standard centralized methods (in this case regulator equation solutions). This solution effectively separates the controller synthesis into two parts -- distributed exosystem generators by network consensus and local output regulation by regulator equation solution.  This two-part structure of distributed controllers is previously proposed to solve a closely-related output synchronization problem \cite{ScaSep:09,WieSepAll:11,KimShiSeo:11}.

One important limitation, however, of the above solution is: in both the exosystem generator design and the regulator equation solution, it is assumed that each agent uses exactly the same dynamic model as the exosystem. This assumption is unreasonable in the distributed network setting, because those agents that cannot measure the reference signal are unlikely to know the {\it precise} dynamic model of the exosystem. To deal with this challenge, \cite{CaiLewHuHua:15} proposes (in the case of static networks) an ``adpative'' exosystem generator and an adaptive solution to the regulator equations. In essence, each agent runs an additional consensus algorithm to update their ``local estimates'' of the exosystem dynamics. 

All the regulator-equation based solutions above fall short in addressing the issue of system parameter uncertainty. In practice we may not have precise knowledge of some entries of the system matrices, or over time the values of some parameters drift. The distributed output regulation problem considering parameter uncertainty is studied in 
 \cite{WanHonHuaJia:10,SuHonHua:13}. The proposed controller is based on the internal model principle, but does not employ the two-part structure mentioned above. It seems to be for this reason that restrictive conditions (acyclic graph or homogeneous nominal agent dynamics) have to be imposed in order to ensure solving output regulation. Moreover, it is also assumed in \cite{WanHonHuaJia:10,SuHonHua:13} that each agent knows the exact model of the exosystem dynamics.  
In addition, \cite{LiuHuang:15} uses the two-part structure and deals with system parameter uncertainty for linear systems; however, only minimum-phase linear systems are considered.
 
In this paper, we further study the distributed output regulation problem of heterogeneous linear systems that are subject to parameter uncertainty and generally non-minimum-phase. In particular,
we propose to use the two-part structure of the distributed controller in the following manner: The first part of exosystem generator extends that in \cite{CaiLewHuHua:15} to work over time-varying networks (see also  \cite{LiuHuang:16}), and the second part is a dynamic compensator embedding an internal model of the exosystem that addresses parameter uncertainty. The challenge here is, in the design of the dynamic compensator, those agents that cannot directly measure the exosystem have no knowledge of the internal model of the exosystem; on the other hand, we know from \cite{FraWon:76} that a precise internal model is crucial to achieve perfect regulation  with uncertain parameters. To deal with this problem, we propose an extra consensus protocol to update the agents' local estimates of the internal model of the exosystem, and present conditions under which the proposed distributed controller solves the distributed output regulation problem. 

The main contribution of this paper is the novel design of the internal model based dynamic compensator, without requiring all agents to know the precise internal model of the exosystem. This dynamic compensator, combined with the exosystem generator extended from \cite{CaiLewHuHua:15}, provides a fully distributed solution to the distributed output regulation problem over time-varying networks and with parameter uncertainty.  Compared to \cite{CaiLewHuHua:15} (and \cite{ScaSep:09,WieSepAll:11,KimShiSeo:11,SuHua:TAC12}), this work deals with parameter uncertainty. On the other hand, compared to  \cite{WanHonHuaJia:10,SuHonHua:13,LiuHuang:15}, this work does not require acyclic graphs or homogeneous agents, nor is minimum-phase required, and most importantly the internal model of the exosystem is not known {\it a priori}. 
Finally, since the agents gradually ``learn'' the internal model of the exosystem in a distributed fashion (based on consensus protocols), we call this design ``distributed internal model principle'' (with reference to the centralized version \cite{FraWon:76}). 

The rest of the paper is organized as follows.
Section~\ref{Sec_ProFor} formulates the robust output regulation problem. Section~\ref{Sec_Controller} presents the solution distributed controller, which consists of two parts -- a distributed exosystem generator and a distributed dynamic compensator.
Section~\ref{Sec_Result} states our main result and provides its proof.
Section~\ref{Sec_Example} illustrates our result by an example. Finally, Section~\ref{Sec_Concl} states our conclusions.


%
\section{Problem Formulation} \label{Sec_ProFor}

Consider a network of $N$ agents, and model their interconnection
structure by a time-varying digraph $\mathcal {G}(t) = (\mathcal {V}, \mathcal
{E}(t))$: Each \emph{node} in $\mathcal {V} = \{ 1,...,N \}$ stands for
an agent, and each directed \emph{edge} $(j,i)$ in $\mathcal {E}(t)
\subseteq \mathcal {V} \times \mathcal {V}$ denotes that agent $j$
communicates to agent $i$ at time $t$ (namely, the information flow is from $j$
to $i$).
In $\mathcal {G}(t)$ a node $i$ is \emph{reachable} from a node $j$ if
there exists a path from $j$ to $i$ which respects the direction of
the edges. We say that $\mathcal {G}(t)$ contains a \emph{globally
reachable} node $i$ if every other node is reachable from $i$
(equivalently $\mathcal {G}(t)$ contains a \emph{spanning tree} with
$i$ the root). For a time interval $[t_1,t_2]$ define the \emph{union
digraph} $\mathcal{G}([t_1,t_2]) := \left( \mathcal {V}, \bigcup_{t
\in [t_1,t_2]} \mathcal {E}(t) \right)$; namely, the edge set of
$\mathcal {G}([t_1,t_2])$ is the union of those over the interval
$[t_1,t_2]$.  We say that $\mathcal {G}(t)$ \emph{uniformly contains a globally reachable node} if there is $T > 0$ such that for every $t_1$ the
union digraph $\mathcal {G}([t_1,t_1+T])$ contains a globally reachable node.

We consider agents that are linear, time-invariant, and
finite-dimensional; namely for each $i \in \mathcal {V}$,
\begin{align}
\dot{x}_i &= A_i x_i + B_i u_i + P_i w_0 \label{eq:state} \\
z_i &= C_i x_i + D_i u_i + Q_i w_0 \label{eq:regulated}
\end{align}
where $x_i \in \mathbb{R}^{n_i}$ is the state vector, $u_i \in
\mathbb{R}^{m_i}$ the control input, and $z_i \in \mathbb{R}^{q_i}$
the output to be regulated. The exogenous signal $w_0 \in
\mathbb{R}^{r}$ satisfies
\begin{align} \label{eq:exosys}
\dot{w}_0 &= S w_0
\end{align}
and represents reference to be tracked and/or disturbance to be
rejected. (\ref{eq:exosys}) is called the \emph{exosystem}. Typically
the vector $P_i w_0$ in (\ref{eq:state}) represents disturbance
acting on the agent dynamics, and $Q_i w_0$ in (\ref{eq:regulated})
represents reference signals to be tracked.

Note that the agents are generally heterogeneous, in that the
entries of the matrices in (\ref{eq:state}) and
(\ref{eq:regulated}), even their dimensions, may be different.
Moreover, we consider that these matrices are uncertain in the
following sense (cf. \cite{FraWon:76}): Viewing $(A_i, B_i, C_i, D_i, P_i, Q_i)$ as a data
point $\wp_i$ (i.e. a vector) in the Euclidean space
$\mathbb{R}^{(n_i+q_i)(n_i+m_i+r)}$, there is a open neighborhood
$\mathcal {P}_i$ of $\wp_i$ in which the entries of the matrices may
vary. The neighborhood need not be small. This uncertainty may 
reflect again heterogeneity of agents, as well as imprecision knowledge 
of system parameters.

In the distributed time-varying network setup, only a few agents may have access to
information of the exosystem (\ref{eq:exosys}) at some time. To formalize this,
denote the exosystem by node ``$0$'' and let $\mathcal {V}_0(t)$ be a
strict subset of $\mathcal {V}$ that have access to node 0 at time $t$.
Namely, the nodes in $\mathcal {V} \setminus \mathcal {V}_0(t)$ do not
have knowledge of the exogenous signal $w_0$, nor do they know 
the exosystem's dynamics $S_0:=S$. 

The extended digraph $\hat{\mathcal {G}}(t)$ 
including node 0 is therefore $\hat{\mathcal {G}}(t) = ( \hat{\mathcal {V}}, \hat{\mathcal
{E}}(t) )$, where $\hat{\mathcal {V}}=\mathcal {V} \cup \{0\}$ and
$\hat{\mathcal {E}}(t)=\mathcal {E}(t) \cup \{(0,i) | i \in \mathcal
{V}_0(t)\}$.  Define the \emph{neighbor set} of $i \in \hat{\mathcal
{V}}$ at time $t$ by $\mathcal {N}_i(t) := \{j \in \hat{\mathcal {V}} \,|\, (j,i) \in
\hat{\mathcal {E}}(t) \}$. Note that $\mathcal {N}_0(t) = \emptyset$ for all $t$, i.e.
the exosystem does not receive information from the rest. Moreover, 
associate a nonnegative weight $a_{ji}(t)$ to each pair $(j,i) \in \hat{\mathcal{V}} \times \hat{\mathcal{V}}$, with $a_{ji}(t) \geq \alpha >0$ for $(j,i) \in \hat{\mathcal {E}}(t)$ and $a_{ji}(t) =0$ for $(j,i) \notin \hat{\mathcal {E}}(t)$, where $\alpha$ is a positive constant. We assume that $a_{ji}(t)$ is piecewise continuous and bounded for all $t \geq 0$.

The graph Laplacian $L(t) \in \mathbb{R}^{(N+1) \times (N+1)}$ of $\hat{\mathcal {G}}(t)$ is given by
\begin{align} \label{eq:laplacian}
l_{ij}(t) := \begin{cases} 
   \sum_{i=0}^N a_{ij}(t), & \text{if } i=j; \\
   -a_{ij}(t),       & \text{if } i \neq j.
  \end{cases}
\end{align}
Also define $L^-(t) \in \mathbb{R}^{N \times N}$ to be the matrix obtained by removing the first row and the first column of $L(t)$ (corresponding to node $0$). It is evident that $L(t)$ and $L^-(t)$ are piecewise continuous and bounded for all $t \geq 0$.


Despite that not all agents can access information of node 0, and
despite uncertainty of the matrices in (\ref{eq:state}) and
(\ref{eq:regulated}), our goal is to design distributed controllers
to achieve {\it output regulation}, which means that $z_i(t) \rightarrow 0$ (uniformly exponentially) as $t
\rightarrow \infty$ for agents $i \in \mathcal {V}$ and all initial conditions $x_i(0)$, $w_0(0)$.

\medskip 

\emph{Problem.} For each agent $i \in \mathcal {V}$, design a
distributed controller such that output regulation holds for some open neighborhood 
$\mathcal {P}_i$ of $\wp_i$.   

\medskip

We will solve this problem in the following sections. Notation: for a complex number $\lambda$, write $Re(\lambda)$ for its real part and $Im(\lambda)$ its imaginary part. Let $\iota := \sqrt{-1}$. 
For two sets ${\cal S}_1$ and ${\cal S}_2$ of finite number of complex numbers, their distance is 
\begin{align*}
d({\cal S}_1, {\cal S}_2) := \min \{ |s_1-s_2| \,|\, s_1 \in {\cal S}_1, s_2 \in {\cal S}_2\}.
\end{align*}


%
\section{Structure of Distributed Controller}
\label{Sec_Controller}

At the outset we make the following standing assumptions, which are either necessary for solving output regulation and consensus or with no loss of generality \cite{FraWon:76,Mor:04,ScaSep:09}.

(A1) For each $i \in \mathcal{V}$, ($A_i, B_i$) is stabilizable.

(A2) For each $i \in \mathcal{V}$, ($C_i, A_i$) is detectable.

(A3) The real parts of the eigenvalues of $S_0$ are all zero.

(A4) The digraph $\hat{\mathcal {G}}(t)$ uniformly contains a globally reachable node and the node is $0$.

%

(A5) For each $i \in \mathcal{V}$ and for each eigenvalue $\lambda$ of $S_0$,
\begin{align} \label{eq:rank_equality}
\mbox{rank}  \begin{bmatrix}
A_i - \lambda I & B_i \\
C_i & D_i
\end{bmatrix} = n_i + q_i.
\end{align}

(A5) states that the eigenvalues of $S_0$ do not coincide with the transmission zeros of agent $i$ 
(for all $i$). 
Let $\sigma(S_0)$ denote the set of eigenvalues of $S_0$, and  
\begin{align*} 
\zeta_i := \left\{ s \in \mathbb{C} \,|\, 
\mbox{rank}  \begin{bmatrix}
A_i - s I & B_i \\
C_i & D_i
\end{bmatrix} < n_i + q_i \right\}
\end{align*}
be the set of transmission zeros of agent $i$. 
Then (A5) means that $\sigma(S_0) \cap \zeta_i = \emptyset$. 
Moreover, let 
\begin{align*} 
\tilde{\zeta}_i := \{ s \in \zeta_i \,|\, Re(s) = 0\}
\end{align*}
be the subset of purely-imaginary transmission zeros. We define the following quantity, which will be used in the design of distributed dynamic compensator.
\begin{align} \label{eq:delta_i}
\delta_i := \begin{cases} 
0 &\mbox{if } \tilde{\zeta}_i = \emptyset \\ 
d(\sigma(S_0),\tilde{\zeta}_i) & \mbox{if } \zeta_i \setminus \tilde{\zeta}_i = \emptyset \\
\min\{ d(\sigma(S_0),\tilde{\zeta}_i), d(\tilde{\zeta}_i,\zeta_i \setminus \tilde{\zeta}_i) \} & \mbox{otherwise.}
\end{cases} 
\end{align}

\medskip

In the following
we describe the structure of our proposed distributed controller, consisting of
two parts: (1) distributed \emph{exosystem generator} and (2)
distributed \emph{dynamic compensator}.

{\bf Distributed exosystem generator.} Since not all agents may get information
from the exosystem, it is reasonable that each agent $i \in \mathcal {V}$ has a
local estimate of the exosystem. Consider
\begin{align} \label{eq:generator}
\dot{w}_i &= S_i(t) w_i + \sum_{j \in \mathcal {N}_i(t)} a_{ij}(t) (w_j -
w_i)
\end{align}
where $S_i$ follows
\begin{align} \label{eq:S_update}
\dot{S}_i &= \sum_{j \in \mathcal {N}_i(t)} a_{ij}(t) (S_j - S_i).
\end{align}
The purpose of (\ref{eq:generator}), (\ref{eq:S_update}) is for each agent $i$ to estimate, using consensus protocols, the dynamics and states of the exosystem; thus we call (\ref{eq:generator}), (\ref{eq:S_update}) the ``exosystem generator''.

For time-invariant networks, it is proved in \cite{CaiLewHuHua:15} that if note $0$ is the globally reachable node, then for each $i \in \mathcal {V}$, $w_i(0)$ and $S_i(0)$
\begin{align*}
\lim_{t \rightarrow \infty} (S_i(t) - S_0) = 0,\ \ \  \lim_{t \rightarrow \infty} (w_i(t) - w_0(t)) = 0.
\end{align*} 
We shall establish the same result but for time-varying networks under assumption (A4).

\medskip

{\bf Distributed dynamic compensator.} Not being able to directly track $w_0$ of the exosystem
(node 0) for all time, each agent tracks its own $w_i$ generated by (\ref{eq:generator}). Here we assume \emph{error feedback}, i.e.
the following is available for feedback
\begin{align} \label{eq:error}
e_i &= C_i x_i + D_i u_i + Q_i w_i \ \in \mathbb{R}^{q_i}.
\end{align}
This is $z_i$ in (\ref{eq:regulated}) with $w_0$ replaced by $w_i$.
Note that it would not be reasonable to assume the availability of
$z_i$ in (\ref{eq:regulated}) for feedback, because (again) not all agents can access
$w_0$. 

With $e_i$ in (\ref{eq:error}) as input and the control $u_i$
as output, consider the following ``dynamic compensator''
\begin{align} \label{eq:compensator}
\dot{\xi}_i &=  E_i(t) \xi_i + F_i(t) e_i \notag \\ 
u_i &= K_i(t) \xi_i.
\end{align}
Our strategy is to use (\ref{eq:compensator}) to achieve $e_i \rightarrow 0$;
assuming that the exosystem generator works effectively so that $w_i \rightarrow w_0$, 
the desired $z_i \rightarrow 0$ will ensue. In the sequenl 
we specify the matrices $E_i$, $F_i$, and $K_i$ in (\ref{eq:compensator}). 

%

Let the minimal polynomial of $S_0$ be
\begin{align*}
s^k + c_{0,1} s^{k-1} + \cdots c_{0,k-1}s + c_{0,k},\ \ \ k \leq r
\end{align*}
and its roots (i.e. the eigenvalues of $S_0$) be $\lambda_{0,1},...,\lambda_{0,k}$.
By (A3) we have $Re(\lambda_{0,1}),...,Re(\lambda_{0,k}) = 0$.
Write $\lambda_0 := [\lambda_{0,1} \cdots\lambda_{0,k}]^\top \in \mathbb{C}^{k}$, $c_0 := [c_{0,1} \cdots c_{0,k}]^\top \in \mathbb{R}^{k}$, and
\begin{align*} 
\mathbb{C}^k_+ := \{s \in \mathbb{C}^k \ |\ Re(s_1),...,Re(s_k) \geq 0 \}.
\end{align*}
Thus $\lambda_0 \in \mathbb{C}^k_+$.

For each agent $i$, let $\lambda_i :=[\lambda_{i,1} \cdots\lambda_{i,k}]^\top\in \mathbb{C}^{k}$ be a local estimate of $\lambda_0$, and  
$c_i := [c_{i,1} \cdots c_{i,k}]^\top \in \mathbb{R}^{k}$ be a local estimate of $c_0$ where 
the entries satisfy
\begin{align} \label{eq:mp_update}
(s-\lambda_{i,1}) \cdots (s-\lambda_{i,k}) = s^k + c_{i,1} s^{k-1} + \cdots c_{i,k-1}s + c_{i,k}.
\end{align} 
Hence each $c_{i,l}$ ($l=1,...,k$) is a polynomial in $\lambda_i =  [\lambda_{i,1} \cdots\lambda_{i,k}]^\top$, and we write $c_{i,l}(\lambda_i)$ henceforth. 

Let
{\small \begin{align} \label{eq:1copy_imp}
G'_i(\lambda_i) &:= \begin{bmatrix}
0 & 1 & \cdots & 0 \\
0 & 0 & \cdots & 0 \\
\vdots & \vdots & \ddots & \vdots \\
0 & 0 & \cdots & 1 \\
-c_{i,k}(\lambda_i) & -c_{i,k-1}(\lambda_i) & \cdots & -c_{i,1}(\lambda_i)
\end{bmatrix} \notag \\
H'_i &:= \begin{bmatrix}
0 \\
0 \\
\vdots \\
0 \\
1
\end{bmatrix}.
\end{align}}
Then the $q_i$-copy internal model is
{\small \begin{align} \label{eq:qcopy_imp}
G_i(\lambda_i) := \underbrace{\begin{bmatrix}
G'_i(\lambda_i) &  &  \\
 & \ddots &  \\
 &  & G'_i(\lambda_i)
\end{bmatrix}}_\text{$q_i$ diagonal blocks},\ 
H_i := \underbrace{\begin{bmatrix}
H'_i &  &  \\
 & \ddots &  \\
 &  & H'_i 
\end{bmatrix}}_\text{$q_i$ diagonal blocks}.
\end{align}}

\begin{lem} \label{lem:stabilizable}
Let (A1) and (A2) hold. Then for every $\lambda_i \in \mathbb{C}^k_+ \setminus \zeta_i$, the following pair of matrices 
\begin{align*}
\left( \begin{bmatrix}
A_i & 0 \\
H_i C_i & G_i(\lambda_i)
\end{bmatrix},\ 
\begin{bmatrix}
B_i \\
H_i D_i
\end{bmatrix} \right)
\end{align*}
is stabilizable.
\end{lem}

{\it Proof.} By $\lambda_i \in \mathbb{C}^k_+ \setminus \zeta_i$, for each $j =1, ..., k$ there holds
\begin{align*}
\mbox{rank}  \begin{bmatrix}
A_i - \lambda_{i,j} I & B_i \\
C_i & D_i
\end{bmatrix} = n_i + q_i.
\end{align*}
Then the conclusion follows immediately from Lemma~1.26 of \cite{Hua04}. \hfill $\square$

Let $\lambda_i \in \mathbb{C}^k_+$. Then by e.g. pole assignment, we derive $[K_{i1}(\lambda_i) \ \ K_{i2}(\lambda_i)]$ such that the matrix
\begin{align*}
\begin{bmatrix}
A_i & 0 \\
H_i C_i & G_i(\lambda_i)
\end{bmatrix} + 
\begin{bmatrix}
B_i \\
H_i D_i
\end{bmatrix} 
\begin{bmatrix}
K_{i1}(\lambda_i) & K_{i2}(\lambda_i)
\end{bmatrix} 
\end{align*}
is {\it stable}, i.e. all the eigenvalues have negative real parts. Note that the entries of $[K_{i1}(\lambda_i) \ K_{i2}(\lambda_i)]$ are in general rational polynomial functions of $\lambda_i$. In addition, when (A2) holds, we choose $L_i$ such that the matrix $A_i - L_i C_i$ is stable.

\medskip

Now we are ready to present the matrices $E_i$, $F_i$, and $K_i$ in the dynamic compensator (\ref{eq:compensator}): 
{\footnotesize \begin{align} \label{eq:E_F_K}
E_i(\lambda_i) &:= \begin{bmatrix}
A_i + (B_i-L_i D_i)K_{i1}(\lambda_i) - L_i C_i & (B_i-L_i D_i)K_{i2}(\lambda_i) \\
0 & G_i(\lambda_i)
\end{bmatrix} \notag \\
F_i &:= \begin{bmatrix}
L_i \\
H_i
\end{bmatrix},\ \ \ K_i(\lambda_i) := [K_{i1}(\lambda_i) \ K_{i2}(\lambda_i)]
\end{align}}

\noindent where $\lambda_i = [\lambda_{i,1} \cdots \lambda_{i,\floor*{\frac{k}{2}}} \ \lambda_{i,\floor*{\frac{k}{2}}+1} \cdots \lambda_{i,k} ]^\top$ follows the update date scheme described below.  For each $l \in [1,\floor*{\frac{k}{2}}]$, $\lambda_{i,l}(t) = \alpha_{i,l}(t) + \iota \beta_{i,l}(t)$ where 
{\small \begin{align}
\dot{\beta}_{i,l}(t) &= \sum_{j \in \mathcal {N}_i(t)} a_{ij}(t) (\beta_{j,l}(t) - \beta_{i,l}(t)) \notag \\
\alpha_{i,l}(t) &= 
\begin{cases} 
0 &\mbox{if } |\beta_{i,l}(t) - \gamma_i(t)| \geq \delta_i \\ 
\sqrt{\delta_i^2 - (\beta_{i,l}(t)-\gamma_i(t))^2} & \mbox{if } |\beta_{i,l}(t) - \gamma_i(t)| < \delta_i;\end{cases} 
\label{eq:eig_update}
\end{align}}

\noindent here $\gamma_i(t)$ is such that $\gamma_i(t) \in \tilde{\zeta}_i$ and $|\beta_{i,l}(t) - \gamma_i(t)| = d(\{\beta_{i,l}(t)\}, \tilde{\zeta}_i)$;  namely $\gamma_i(t)$ is the closest (purely-imaginary) transmission zero to $\beta_{i,l}(t)$ at time $t$.  (If $\tilde{\zeta}_i = \emptyset$, then $\delta_i=0$ by (\ref{eq:delta_i}) and thus simply let $\alpha_{i,l}(t) = 0$ for all $t$.)
Set the initial conditions for (\ref{eq:eig_update}) to be $\alpha_{i,l}(0) = 0$ and $\beta_{i,l}(0) \in {\Bbb R}$ such that $d(\{\beta_{i,j}(0)\}, \tilde{\zeta}_i) \geq \delta_i$, for all $l \in [1,\floor*{\frac{k}{2}}]$.  In addition, let  
\begin{align*}
\lambda_{i,\floor*{\frac{k}{2}}+l} (t) = \alpha_{i,l}(t) - \iota \beta_{i,l}(t) 
\end{align*}
for each $l \in [1,\floor*{\frac{k}{2}}]$. Finally set $\lambda_{i,k} = 0$ if and only if $k$ is an odd number. 

Specified as above, the components of $\lambda_i(t)$ are complex numbers symmetric with respect to the real axis for all $t$. Moreover, by (\ref{eq:eig_update}) and assumption (A4), $\beta_{i,l}(t) \rightarrow Im(\lambda_{0,l})$ and $\alpha_{i,l}(t) \rightarrow 0 = Re(\lambda_{0,l})$  (as $t \rightarrow \infty$) for all $l \in [1,\floor*{\frac{k}{2}}]$ \cite{Mor:04}. Also note that 
\begin{align*}
(\forall t \geq 0)\, \alpha_i(t) \geq 0
\end{align*}
This implies that $\lambda_i(t) \rightarrow \lambda_0$ as $t \rightarrow \infty$, and $\lambda_i \in \mathbb{C}^k_+ \setminus \zeta_i$ for all $t \geq 0$. Hence the pair of matrices in Lemma~\ref{lem:stabilizable} is stabilizable for all $t \geq 0$.

Note that in (\ref{eq:E_F_K}), $E_i$ and $K_i$ are time-varying (as $\lambda_i$ is time-varying), while $F_i$ is time-invariant. 


%
\section{Main Result}
\label{Sec_Result}

Our main result is the following.

\begin{thm} \label{thm:main}
Given the multi-agent system (\ref{eq:state}), (\ref{eq:regulated}) and the exosystem (\ref{eq:exosys}), let (A1)-(A5) hold. Then for each agent $i \in \mathcal{V}$, the distributed exosystem generator (\ref{eq:generator}) with (\ref{eq:S_update}) and the distributed dynamic compensator (\ref{eq:compensator}) with (\ref{eq:E_F_K}), (\ref{eq:eig_update}) achieve output regulation, i.e. for all $x_i(0)$ and $w_0(0)$,
\begin{align*}
z_i(t) \rightarrow 0 \mbox{ (uniformly exponentially) as } t \rightarrow \infty
\end{align*}
for some open neighborhood  $\mathcal {P}_i$ of $\wp_i$.
\end{thm}

\medskip

Before proving Theorem~\ref{thm:main}, we state a few remarks concerning this result.

{\it Remark 1:}  For the distributed output regulation problem, Theorem~\ref{thm:main} extends previous results in the literature in several aspects: The proposed distributed controller (i) works effectively over time-varying digraphs (cf. \cite{WanHonHuaJia:10,SuHonHua:13,CaiLewHuHua:15}), (ii) employs internal model to deal with system parameter uncertainty (cf. \cite{XiaWeiLi:09,SuHua:TAC12,CaiLewHuHua:15}), (iii) needs no {\it a priori} knowledge of the exosystem (cf. \cite{XiaWeiLi:09,WanHonHuaJia:10,SuHua:TAC12,SuHonHua:13}), and (iv) deals with generally non-minimum-phase systems (cf. \cite{LiuHuang:15}).

{\it Remark 2:} For the distributed synchronization problem, our distributed controller may be applied by treating the exosystem as the ``virtual leader'', thereby improving the solutions of \cite{ScaSep:09,WieSepAll:11,KimShiSeo:11} to deal with disturbance rejection, parameter uncertainty, and initially unknown internal model.

{\it Remark 3:} For each agent to ``learn'' the internal model of the exosystem, our strategy is to make the agents reach consensus by (\ref{eq:eig_update}) at the {\it eigenvalues} of the exosystem's minimal polynomial. It might appear more straightforward to reach consensus at the {\it coefficients} of the exosystem's minimal polynomial; the advantage of updating $\lambda_i$ with (\ref{eq:eig_update}), however, is that we may directly guarantee the equality in (\ref{eq:rank_equality}) in assumption (A5).

If the exosystem is a leader agent that possesses computation and communication abilities, then the leader can compute the eigenvalues of its own minimal polynomial and send the information to other agents. If the exosystem is some entity that cannot compute or communicate, then those agents that can measure the exosystem (in particular know $S_0$) compute the corresponding minimal polynomial and the eigenvalues, and send the information to the rest of the network.


\medskip

In the following, we prove Theorem~\ref{thm:main}. First we need two lemmas, which establishes that the distributed exosystem generator (\ref{eq:generator}) with (\ref{eq:S_update}) works effectively for time-varying networks. The proofs are in Appendix.\footnote{It has been brought to our attention recently that these results are derived independently in \cite{LiuHuang:16}, which is reaffirmation of their correctness.  Nevertheless the main novelty of our approach lies in the design of the distributed dynamic compensator, in particular the update scheme of the local eigenvalue estimates; the latter contributes to solving the problem for generally non-minimum-phase agents.}

\begin{lem} \label{lem:A1A2A3}
Consider
\begin{align}  \label{eq:A1A2A3}
\dot{x} = A_1(t) x(t) + A_2(t) x(t) + A_3(t)
\end{align}
where $A_1(t)$, $A_2(t)$, $A_3(t)$ are piecewise continuous and bounded on $[0,\infty)$. Suppose that the origin is a uniformly exponentially stable equilibrium of $\dot{x} = A_1(t) x$, and $A_2(t) \rightarrow 0$, $A_3(t) \rightarrow 0$ (uniformly exponentially) as $t \rightarrow \infty$. Then $x(t) \rightarrow 0$ (uniformly exponentially) as $t \rightarrow \infty$.
\end{lem}



\begin{lem} \label{lem:generator}
Consider the distributed exosystem generator (\ref{eq:generator}), (\ref{eq:S_update}). If (A4) holds, then for each $i \in \mathcal {V}$, $S_i(0)$ and $w_i(0)$, there holds $S_i(t) \rightarrow S_0$ and $w_i(t) \rightarrow w_0(t)$ (uniformly exponentially) as $t \rightarrow \infty$.
\end{lem}

 \medskip

We are ready to prove Theorem~\ref{thm:main}.

{\it Proof.} Suppose that (A1)-(A5) hold.
Fix an agent $i \in \mathcal{V}$, and consider the combined state $\mu_i := [x^\top_i \ \xi^\top_i]^\top$ of the agent and its dynamic compensator. From (\ref{eq:state}), (\ref{eq:regulated}), (\ref{eq:compensator}) we derive
\begin{align*}
\begin{bmatrix}
\dot{x}_i \\
\dot{\xi}_i
\end{bmatrix} &=
\begin{bmatrix}
A_i & B_i K_i(\lambda_i(t)) \\
F_i C_i & E_i(\lambda_i(t)) + F_i D_i K_i(\lambda_i(t))
\end{bmatrix} 
\begin{bmatrix}
x_i \\
\xi_i
\end{bmatrix}\\  
&\ \ \ \ + 
\begin{bmatrix}
P_i \\
F_i Q_i
\end{bmatrix} w_i  +
\begin{bmatrix}
P_i \\
0
\end{bmatrix} (w_0-w_i) \\
z_i &= \begin{bmatrix}
C_i & D_i K_i(\lambda_i(t))
\end{bmatrix} 
\begin{bmatrix}
x_i \\
\xi_i
\end{bmatrix}  + 
Q_i w_0
\end{align*}

Write
\begin{align*}
M_i(\lambda_i(t)) := 
\begin{bmatrix}
A_i & B_i K_i(\lambda_i(t)) \\
F_i C_i & E_i(\lambda_i(t)) + F_i D_i K_i(\lambda_i(t))
\end{bmatrix}. 
\end{align*}
By (A4) and the update (\ref{eq:eig_update}) with the initial condition $\lambda_i(0) \in \mathbb{C}^k_+ \setminus \zeta_i$ for all $i$, it is derived that
\begin{align*}
\lambda_i(t) \rightarrow \lambda_0 \mbox{ (uniformly exponentially) as } t \rightarrow \infty
\end{align*}
and moreover
\begin{align*}
(\forall t \geq 0)\ \lambda_i(t) \in \mathbb{C}^k_+ \setminus \zeta_i.
\end{align*}
Hence by Lemma~\ref{lem:stabilizable}, for every $t \geq 0$ the pair
\begin{align*}
\left( \begin{bmatrix}
A_i & 0 \\
H_i C_i & G_i(\lambda_i(t))
\end{bmatrix},\ 
\begin{bmatrix}
B_i \\
H_i D_i
\end{bmatrix} \right)
\end{align*}
is stabilizable, and the stabilizing gain matrix 
$K_i(\lambda_i(t)) = \begin{bmatrix}
K_{i1}(\lambda_i(t)) & K_{i2}(\lambda_i(t))
\end{bmatrix}$
is well-defined.
It follows from $\lambda_i(t) \rightarrow \lambda_0$ (uniformly exponentially) that
\begin{align*}
& K_i(\lambda_i(t)) \rightarrow K_i(\lambda_0) \\
& G_i(\lambda_i(t)) \rightarrow G_i(\lambda_0) \\
& E_i(\lambda_i(t)) \rightarrow E_i(\lambda_0)
\end{align*}
and therefore
\begin{align*}
M_i(\lambda_i(t)) \rightarrow M_i(\lambda_0) \mbox{ (uniformly exponentially).}
\end{align*}

Let us analyze $M_i(\lambda_0)$. By (\ref{eq:E_F_K}) we have
{\footnotesize \begin{align*}
&M_i(\lambda_0) = 
\begin{bmatrix}
A_i & B_i K_i(\lambda_0) \\
F_i C_i & E_i(\lambda_0) + F_i D_i K_i(\lambda_0)
\end{bmatrix}  \\
&= \begin{bmatrix}
A_i & B_i K_{i1}(\lambda_0) & B_i K_{i2}(\lambda_0) \\
L_i C_i & A_i+B_i K_{i1}(\lambda_0)-L_i C_i & B_i K_{i2}(\lambda_0) \\
H_i C_i & H_i D_i K_{i1}(\lambda_0) & G_i(\lambda_0)+H_i D_i K_{i2}(\lambda_0)
\end{bmatrix} 
\end{align*}}
Subtracting the first row from the second row, and then adding the second column to the first column yield 
{\footnotesize \begin{align*}
\begin{bmatrix}
A_i+B_i K_{i1}(\lambda_0) & B_i K_{i1}(\lambda_0) & B_i K_{i2}(\lambda_0) \\
0 & A_i-L_i C_i & 0 \\
H_i C_i + H_i D_i K_{i1}(\lambda_0) & H_i D_i K_{i1}(\lambda_0) & G_i(\lambda_0)+H_i D_i K_{i2}(\lambda_0)
\end{bmatrix} 
\end{align*}}
By (A5), we have $\lambda_0 \in \mathbb{C}^k_+ \setminus \zeta_i$, and again by Lemma~\ref{lem:stabilizable} the pair
\begin{align*}
\left( \begin{bmatrix}
A_i & 0 \\
H_i C_i & G_i(\lambda_0)
\end{bmatrix},\ 
\begin{bmatrix}
B_i \\
H_i D_i
\end{bmatrix} \right)
\end{align*}
is stabilizable, and $K_i(\lambda_0) = \begin{bmatrix}
K_{i1}(\lambda_0) & K_{i2}(\lambda_0)
\end{bmatrix}$ is such that 
\begin{align*}
\begin{bmatrix}
A_i + B_i K_{i1}(\lambda_0) & B_i K_{i2}(\lambda_0) \\
H_i C_i + H_i D_i K_{i1}(\lambda_0) & G_i(\lambda_0) + H_i D_i K_{i2}(\lambda_0) 
\end{bmatrix}
\end{align*}
is stable. Moreover, $L_i$ is chosen such that $A_i - L_i C_i$ is stable. 
Therefore $M_i(\lambda_0)$ is stable. 

Since eigenvalues of a matrix are continuous functions of the entries of the matrix, there exists a neighborhood $\mathcal{P}_i$ of the data point $\wp_i$ such that $M_i(\lambda_0)$ remains stable.
Thus for every data point in $\mathcal{P}_i$, the following equations
\begin{align*}
X_i(\lambda_0) S &= M_i(\lambda_0)X_i(\lambda_0) + \begin{bmatrix}
P_i \\
F_i Q_i
\end{bmatrix} \\
0 &= \begin{bmatrix}
C_i & D_i K_i(\lambda_0)
\end{bmatrix} 
X_i(\lambda_0) + Q_i
\end{align*}
have a unique solution $X_i(\lambda_0)$.

Let $\tilde{\mu}_i := \mu_i - X_i(\lambda_0)w_i$ and $\tilde{M}_i(t) := M_i(\lambda_i(t)) - M_i(\lambda_0)$. 
Then
{\small \begin{align*}
\dot{\tilde{\mu}}_i &= \dot{\mu}_i - X_i(\lambda_0) \dot{w}_i \\
&= M_i(\lambda_i(t)) \mu_i + \begin{bmatrix}
P_i \\
F_i Q_i
\end{bmatrix} w_i +
\begin{bmatrix}
P_i \\
0
\end{bmatrix} (w_0-w_i) - X_i(\lambda_0) S w_i \\
&= (\tilde{M}_i(t) + M_i(\lambda_0))(\tilde{\mu}_i + X_i(\lambda_0)w_i) + \begin{bmatrix}
P_i \\
F_i Q_i
\end{bmatrix} w_i\\
& \hspace{3.9cm} +
\begin{bmatrix}
P_i \\
0
\end{bmatrix} (w_0-w_i) - X_i(\lambda_0) S w_i \\
&\hspace{-0.5cm} = M_i(\lambda_0) \tilde{\mu}_i + \tilde{M}_i(t) \tilde{\mu}_i + (M_i(\lambda_0)X_i(\lambda_0) + \begin{bmatrix}
P_i \\
F_i Q_i
\end{bmatrix}-X_i(\lambda_0) S) w_i \\
& \hspace{3.3cm} + \tilde{M}_i(t) X_i(\lambda_0) w_i + \begin{bmatrix}
P_i \\
0
\end{bmatrix} (w_0-w_i) \\
&= M_i(\lambda_0) \tilde{\mu}_i + \tilde{M}_i(t) \tilde{\mu}_i + \tilde{M}_i(t) X_i(\lambda_0) w_i + \begin{bmatrix}
P_i \\
0
\end{bmatrix} (w_0-w_i)
\end{align*}}
Since $M_i(\lambda_0)$ is stable for every data point in $\mathcal{P}_i$, $\tilde{M}_i(t) \rightarrow 0$ and $w_0(t)-w_i(t) \rightarrow 0$ (uniformly exponentially) as $t \rightarrow \infty$, it follows from Lemma~\ref{lem:A1A2A3} that $\tilde{\mu}_i(t) \rightarrow 0$ (uniformly exponentially) as $t \rightarrow \infty$ for every data point in $\mathcal{P}_i$.

Finally, the regulated variable $z_i$ is
{\small \begin{align*}
z_i &= C_i x_i + D_i u_i + Q_i w_0 \\
&= C_i x_i + D_i K_i(\lambda_i(t)) \xi_i +Q_i w_i + Q_i(w_0-w_i) \\
&= \begin{bmatrix}
C_i & D_i K_i(\lambda_i(t))
\end{bmatrix} \mu_i + Q_i w_i + Q_i(w_0-w_i) \\
&= \begin{bmatrix}
C_i & D_i K_i(\lambda_i(t))
\end{bmatrix} (\tilde{\mu}_i+X_i(\lambda_0)w_i) + Q_i w_i \\
& \hspace{5.4cm} + Q_i(w_0-w_i) \\
&\hspace{-0.5cm} = \begin{bmatrix}
C_i & D_i K_i(\lambda_i(t))
\end{bmatrix} \tilde{\mu}_i + \left(\begin{bmatrix}
C_i & D_i K_i(\lambda_i(t))
\end{bmatrix}X_i(\lambda_0) + Q_i \right) w_i\\ 
& \hspace{5.4cm} + Q_i(w_0-w_i)
\end{align*}}
Since
\begin{align*}
&\tilde{\mu}_i(t) \rightarrow 0 \\
&\begin{bmatrix}
C_i & D_i K_i(\lambda_i(t))
\end{bmatrix}X_i(\lambda_0) + Q_i  \\
&\hspace{0.85cm} \rightarrow \begin{bmatrix}
C_i & D_i K_i(\lambda_0)
\end{bmatrix}X_i(\lambda_0) + Q_i =0 \\
&w_i(t)) \rightarrow w_0(t)
\end{align*}
(uniformly exponentially) as $t \rightarrow \infty$,
we conclude that 
\begin{align*}
z_i(t) \rightarrow 0 \mbox{ (uniformly exponentially) as } t \rightarrow \infty
\end{align*}
for every data point in $\mathcal{P}_i$. \hfill $\square$

\section{Simulation Example} \label{Sec_Example}

\begin{figure}[!t]
  \centering
  \includegraphics[width=0.45\textwidth]{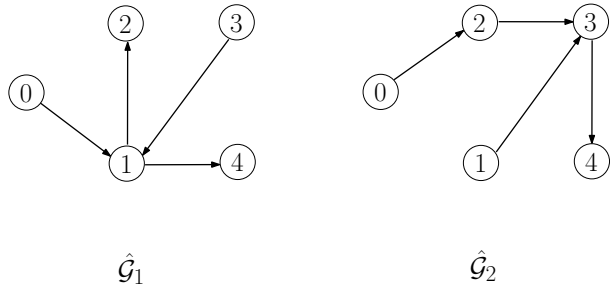}
  \caption{Time-varying network that periodically switches between $\hat{\mathcal{G}}_1$ and $\hat{\mathcal{G}}_2$. Node $0$ is the exosystem, and nodes $1,2,3,4$ are agents.}
  \label{fig:net_top}
\end{figure}

In this section, we illustrate the designed distributed controller by applying it to solve a distributed output regulation problem. As displayed in Fig.~\ref{fig:net_top}, consider a network of 4 agents (nodes 1,2,3.4) and an exosystem (node 0), with two possible topologies $\hat{\mathcal{G}}_1$ and $\hat{\mathcal{G}}_2$. The network is made time-varying by periodic switching between $\hat{\mathcal{G}}_1$ and $\hat{\mathcal{G}}_2$ for equal length of time. Note that neither $\hat{\mathcal{G}}_1$ nor $\hat{\mathcal{G}}_2$ contains a globally reachable node, but their union does and the node is $0$. Owing to the periodic switching, the network uniformly contains the globally reachable node $0$, i.e. assumption (A4) holds. 

The exosystem (node 0) is
\begin{align*}
\dot{w}_0 &= S_0 w_0,\ \ \ S_0 =  \begin{bmatrix}
0 & 2 \\
-2 & 0
\end{bmatrix}.
\end{align*}
The agents ($i =1,2,3,4$) are
\begin{align*}
\dot{x}_i &= A_i x_i + B_i u_i + P_i w_0 \\
z_i &= C_i x_i + D_i u_i + Q_i w_0
\end{align*}
where
\begin{align*}
&A_1 =  \begin{bmatrix}
0 & 1.6 \\
-1.6 & 0
\end{bmatrix},\ A_2 =  \begin{bmatrix}
0 & 1.7 \\
-1.7 & 0
\end{bmatrix} \\
&A_3 =  \begin{bmatrix}
0 & 1.8 \\
-1.8 & 0
\end{bmatrix},\ A_4 =  \begin{bmatrix}
0 & 2.5 \\
-2.5 & 0
\end{bmatrix} \\
& B1 = \cdots B_4 =  \begin{bmatrix}
0  \\
1
\end{bmatrix},\ P_1 = \cdots = P_4 = 0 \\
& C1 = \cdots C_4 =  \begin{bmatrix}
1 & 0
\end{bmatrix},\ D_1 = \cdots = D_4 = 0 \\
& Q1 = \cdots Q_4 =  \begin{bmatrix}
-1 & 0
\end{bmatrix}.
\end{align*}
Moreover, consider that the matrices $A_i$ ($i=1,2,3,4$) are being perturbed as follows:
\begin{align*}
A_i +
\begin{bmatrix}
0 & 0.1 \\
-0.1 & 0
\end{bmatrix}.
\end{align*}
The perturbation term is unknown and thus reflects parameter uncertainty. Let the initial conditions $w_0(0)$ and $x_i(0)$ be chosen uniformly at random from $[-1,1]$. The goal of output regulation is to achieve $\lim_{t \rightarrow \infty} z_i(t) = 0$ for all $i=1,2,3,4$.

\begin{figure}[!t]
  \centering
  \includegraphics[width=0.53\textwidth]{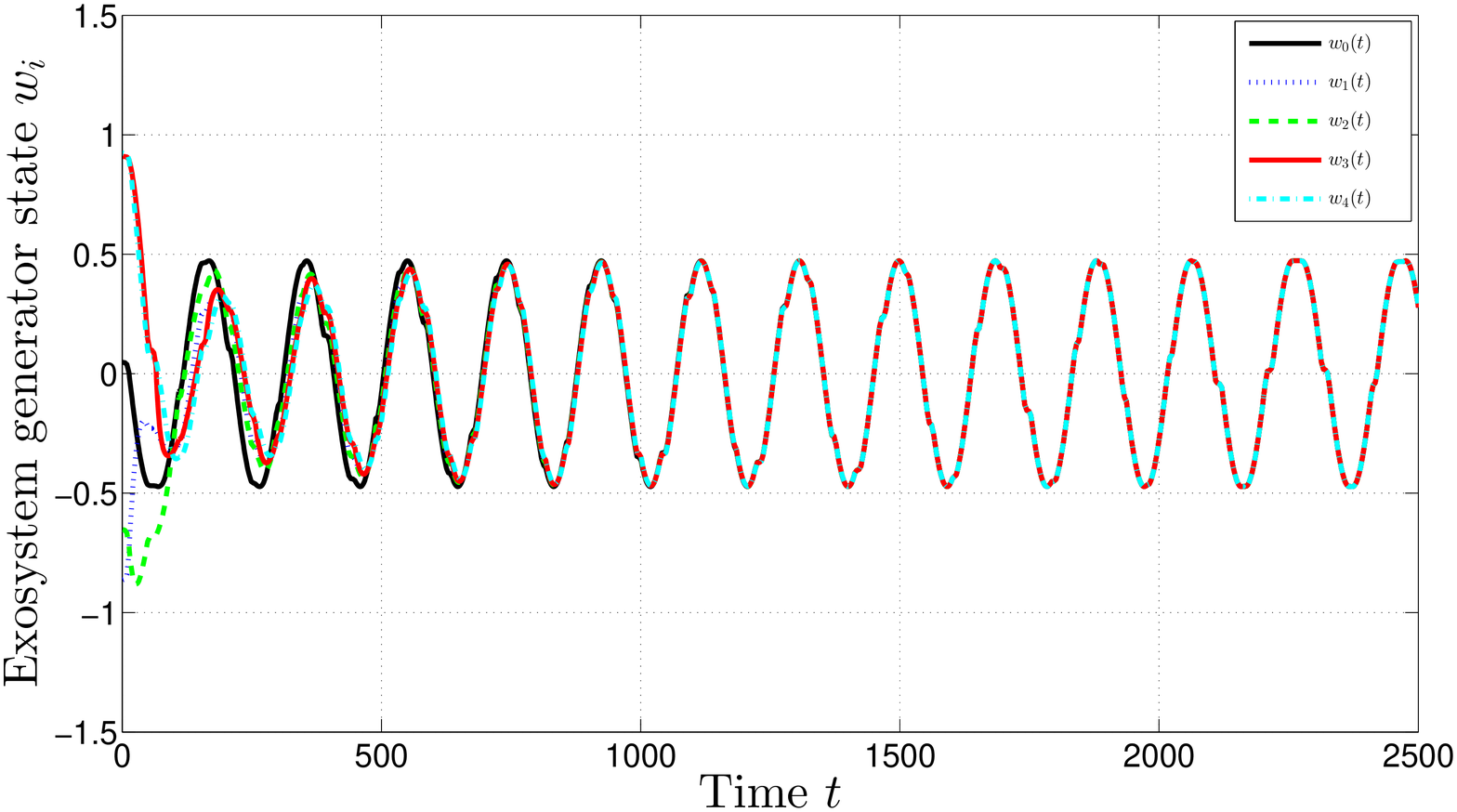}
  \caption{States $w_i(t)$ ($i=1,...4$) of exosystem generators synchronize with the exosystem's signal $w_0(t)$.}
  \label{fig:exo_gen}
\end{figure}

We apply the proposed distributed controller in Section~III and show simulation results below. 
First, apply the distributed exosystem generators (\ref{eq:generator}), (\ref{eq:S_update})
with the initial conditions $w_i(0)$ selected uniformly at random from $[-1,1]$ and $S_i(0)=A_i$ ($i=1,...,4$). The result is displayed in Fig.~\ref{fig:exo_gen}: All $w_i(t)$ ($i=1,...,4$) synchronize with the exosystem's signal $w_0(t)$. Thus the distributed exosystem generators effectively create a local copy of the exosystem, despite that not all agents have access to the exosystem and the network is time-varying.

\begin{figure}[!t]
  \centering
  \includegraphics[width=0.53\textwidth]{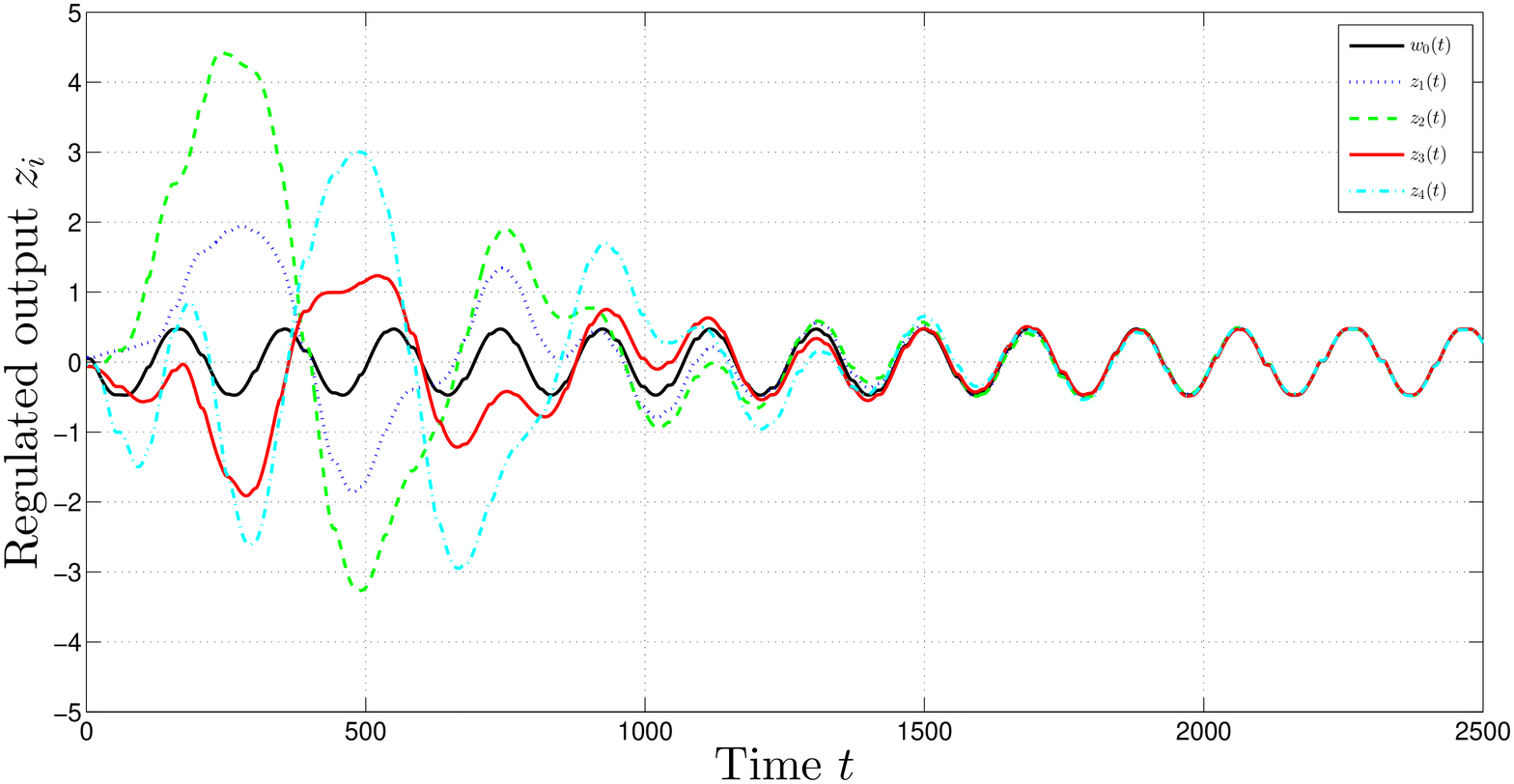}
  \caption{Regulated output variables $z_i(t)$ ($i=1,...4$) converge to the exosystem's signal $w_0(t)$.}
  \label{fig:regulated_output}
\end{figure}

Next, apply the distributed dynamic compensators (\ref{eq:compensator}), (\ref{eq:E_F_K}), (\ref{eq:eig_update}) with initial conditions $\xi_i(0)$, $Im(\lambda_i(0))$ selected uniformly at random from $[-1,1]$ and $Re(\lambda_i(0))$ from $[0,1]$. The result is displayed in Fig.~\ref{fig:regulated_output}: All regulated outputs $z_i(t)$ ($i=1,...,4$) converge to the exosystem's signal $w_0(t)$. This demonstrates the effectiveness of the distributed dynamic compensators for achieving perfect regulation, despite of the parameter perturbation and initially imprecise internal model of the exosystem.

%
\section{Conclusions} \label{Sec_Concl}

We have studied a multi-agent output regulation problem, where the (linear) agents are heterogeneous, subject to parameter uncertainty, and the network is time-varying. The challenge is that the exosystem's dynamics is not accessible by all agents, and consequently the agents do not initially possess a precise internal model of the exosystem. We have solved the problem by proposing a distributed controller consisting of two parts -- an exosystem generator that ``learns'' the dynamics of the exosystem and a dynamic compensator that ``learns'' the internal model. The effectiveness of this solution suggests a {\it distributed internal model principle}: converging internal models imply network output regulation. 



%


\bibliographystyle{IEEEtran}
\bibliography{DistributedControl}

%
%
%
%
%
%

\appendix
{\it Proof of Lemma~\ref{lem:A1A2A3}.} Since the origin is a uniformly exponentially stable equilibrium of $\dot{x} = A_1(t) x$, there exist bounded and positive definite matrices $P_1(t), Q_1(t)$ (for all $t \geq 0$) such that 
\begin{align*}
\dot{P}_1(t) + P_1(t) A_1(t) + A_1(t)^\top P_1(t) = -Q_1(t).
\end{align*}
Then $V_1(x,t) := x^\top P_1(t) x$ is a quadratic Lyapunov function for $\dot{x} = A_1(t) x(t)$, and there exist constants $c_1,c_2,c_3,c_4$ such that the following are satisfied (globally):
\begin{align*}
& c_1 ||x||^2 \leq V_1(x,t) \leq c_2 ||x||^2 \\
& \frac{\partial V_1}{\partial t} + \frac{\partial V_1}{\partial x} A_1(t) x \leq -c_3 ||x||^2 \\
& ||\frac{\partial V_1}{\partial x}|| \leq c_4 ||x||.
\end{align*}
Now consider $\dot{x} = A_1(t)x + A_2(t)x$. The term $A_2(t)x$ satisfies the inequality
\begin{align*}
|| A_2(t)x || \leq || A_2(t) || \cdot ||x||.
\end{align*}
Since $A_2(t) \rightarrow 0$, we have $|| A_2(t) || \rightarrow 0$.
Hence viewing $A_2(t)x$ as a vanishing perturbation to $\dot{x} = A_1(t)x$, it follows from Corollary~9.1 and Lemma~9.5 of \cite{Kha:02} that the origin is also a uniformly exponentially stable equilibrium of $\dot{x} = A_1(t)x + A_2(t)x$. In turn, there exist bounded and positive definite matrices $P_2(t), Q_2(t)$ (for all $t \geq 0$) such that \begin{align*}
\dot{P}_2(t) &+ P_2(t) (A_1(t)+A_2(t)) \\
&+ (A_1(t)+A_2(t))^\top P_2(t) = -Q_2(t).
\end{align*}
Let $V_2(x,t) := x^\top P_2(t) x$ be a candidate Lyapunov function for (\ref{eq:A1A2A3}). Then 
\begin{align*}
\frac{\partial V_2}{\partial t} &+ \frac{\partial V_2}{\partial x} ( A_1(t)x + A_2(t)x + A_3(t) ) \\
&= -x^\top Q_2(t) x + 2 x^\top P_2(t) A_3(t) x\\
&\leq -(||Q_2(t)||-\frac{1}{\epsilon})||x||^2 + \epsilon ||P_2(t)A_3(t)||^2 \\
&\leq -(||Q_2(t)||-\frac{1}{\epsilon})||x||^2 + \epsilon ||P_2(t)||^2 ||A_3(t)||^2.
\end{align*}
Let $\epsilon$ be such that $\epsilon >0$ and $||Q_2(t)||-\frac{1}{\epsilon} >0$. Then it follows from Theorem~5 of \cite{Son:07} that (\ref{eq:A1A2A3}) is input-to-state stable, with $A_3(t)$ the input. Since $A_3(t) \rightarrow 0$ (uniformly exponentially), as a consequence of input-to-state stability (\cite[Section~3.1]{Son:07}, \cite[Section~4.9]{Kha:02}) we conclude that $x(t) \rightarrow 0$ (uniformly exponentially) as $t \rightarrow \infty$.  \hfill $\square$

\bigskip

{\it Proof of Lemma~\ref{lem:generator}.} By (\ref{eq:S_update}) and (A4), for all $S_i(0)$, $S_i(t)$ reach consensus (element wise) uniformly exponentially \cite{Mor:04}. Since node $0$ is the globally reachable node (uniformly), the consensus value is $S_0$, i.e. $S_i(t) \rightarrow S_0$ (uniformly exponentially) as $t \rightarrow \infty$.

To show $w_i(t) \rightarrow w_0(t)$, first consider 
\begin{align} \label{eq:homo-gene}
\dot{w}_i = S_0 w_i + \sum_{j \in \mathcal {N}_i(t)} a_{ij}(t) (w_j -
w_i).
\end{align}
By (A4), for all $w_i(0)$, $w_i(t) \rightarrow w_0$ (uniformly exponentially) as $t \rightarrow \infty$ \cite{ScaSep:09}. Let $\tilde{w}_i := w_i - w_0$. Then
\begin{align*}
\dot{\tilde{w}}_i &= S_0 w_i - S_0 w_0 + \sum_{j \in \mathcal {N}_i(t)} a_{ij}(t) ((w_j - w_0) -
(w_i-w_0)) \\
&= S_0 \tilde{w}_i + \sum_{j \in \mathcal {N}_i(t)} a_{ij}(t) (\tilde{w}_j - \tilde{w}_i).
\end{align*}
Let $\tilde{w} := [\tilde{w}_1 \ \cdots \ \tilde{w}_N]^\top$ and recall $L^-(t)$ defined by removing the first row and the first column of the graph Laplacian $L(t)$ in (\ref{eq:laplacian}). Then
\begin{align} \label{eq:homo-compact}
\dot{\tilde{w}} = (I_N \otimes S_0 - L^-(t) \otimes I_{q}) \tilde{w}.
\end{align}
Since $\tilde{w} \rightarrow 0$ (uniformly exponentially) as $t \rightarrow \infty$, the origin is a uniformly exponentially stable equilibrium of (\ref{eq:homo-compact}). 

Returning to (\ref{eq:generator}) and letting $\tilde{S}_i := S_i - S_0$, we derive
\begin{align*}
\dot{\tilde{w}}_i = S_0 \tilde{w}_i + \tilde{S}_i(t)\tilde{w}_i + \tilde{S}_i(t)w_0 + \sum_{j \in \mathcal {N}_i(t)} a_{ij}(t) (\tilde{w}_j - \tilde{w}_i).
\end{align*}
Hence 
\begin{align*} 
\dot{\tilde{w}} = (I_N \otimes S_0 - L^-(t) \otimes I_{q}) \tilde{w} &+ \mbox{diag}(\tilde{S}_1, \ \cdots, \tilde{S}_N) \tilde{w}\\ 
&+ \mbox{diag}(\tilde{S}_1, \ \cdots, \tilde{S}_N) w_0 {\bf 1}.
\end{align*}
Observe that (i) the matrices $(I_N \otimes S_0 - L^-(t) \otimes I_{q})$, $\mbox{diag}(\tilde{S}_1, \ \cdots, \tilde{S}_N)$, $\mbox{diag}(\tilde{S}_1, \ \cdots, \tilde{S}_N) w_0 {\bf 1}$ are bounded and piecewise continuous; (ii) $\mbox{diag}(\tilde{S}_1, \ \cdots, \tilde{S}_N) \rightarrow 0$, $\mbox{diag}(\tilde{S}_1, \ \cdots, \tilde{S}_N) w_0 {\bf 1} \rightarrow 0$ (uniformly exponentially) as $t \rightarrow \infty$.  Since the origin is a uniformly exponentially stable equilibrium of (\ref{eq:homo-compact}), applying Lemma~\ref{lem:A1A2A3} we conclude that $\tilde{w} \rightarrow 0$ (uniformly exponentially) as $t \rightarrow \infty$. That is, for all $i \in \mathcal{V}$, $w_i(t) \rightarrow w_0(t)$ (uniformly exponentially) as $t \rightarrow \infty$, and the proof is complete.
 \hfill $\square$

\end{document}